\documentclass[pra,twocolumn,nofootinbib,showpacs]{revtex4-1}
\usepackage{amsmath}
\usepackage{amssymb}
\usepackage{bm}
\usepackage[usenames,dvipsnames]{color} %for color
\usepackage{cancel}
\usepackage{dcolumn}
\usepackage{epsf}
\usepackage{epsfig}
\usepackage{graphicx}
\usepackage{mathtools}
\usepackage{multirow}
\usepackage[final]{pdfpages}
\usepackage{setspace}
\usepackage{ulem} %for strikeout
\usepackage{wrapfig}

\definecolor{LinkColor}{rgb}{0,0,.5}
\usepackage[colorlinks=true,linkcolor=BrickRed,
citecolor=LinkColor ,urlcolor=LinkColor]{hyperref}

\newcommand{\half}{\frac{1}{2}}
\newcommand{\beq}{\begin{equation}}
\newcommand{\eeq}{\end{equation}}
                  
\newcommand{\benum}{\begin{enumerate}}
\newcommand{\eenum}{\end{enumerate}}
                    
\newcommand{\bit}{\begin{itemize}}
\newcommand{\eit}{\end{itemize}}

\newcommand{\bea}{\begin{eqnarray}}
\newcommand{\eea}{\end{eqnarray}}

\newcommand{\labs}{\left|}
\newcommand{\rabs}{\right|}

\newcommand{\zr}[1]{Eq.~(\ref{eqn:#1})}

\newcommand{\zfr}[1]{Fig.~\ref{fig:#1}}

\newcommand{\ket}[1]{\left\vert{#1}\right\rangle}
\newcommand{\bra}[1]{\left\langle{#1}\right\vert}
\newcommand{\braket}[2]{\langle{#1}\vert{#2}\rangle}
\newcommand{\ketbra}[2]{\ket{#1}\bra{#2}}
\newcommand{\sand}[3]{\langle{#1}\vert#2\vert{#3}\rangle}

\newcommand{\ba}{\left\{ \begin{array}{lr}}
\newcommand{\ea}{\end{array}\right.}

\newcommand{\blist}[1]{
 \begin{list}{#1}%$\ast\circ\bullet\Rightarrow$\checkmark\star
  { \setlength{\itemsep}{3pt}
     \setlength{\parsep}{2pt}
     \setlength{\topsep}{3pt}
     \setlength{\partopsep}{0pt}
     \setlength{\leftmargin}{1em}
     \setlength{\labelwidth}{1em}
     \setlength{\labelsep}{0.5em} } }
\newcommand{\elist}{
  \end{list}  }

\DeclareMathSymbol{\vartheta}{\mathalpha}{letters}{"12}
\DeclareMathSymbol{\theta}{\mathalpha}{letters}{"23}
\DeclareMathSymbol{\phi}{\mathalpha}{letters}{"27}
\DeclareMathSymbol{\varphi}{\mathalpha}{letters}{"1E}

\begin{document}
\title{Perfect quantum transport in arbitrary spin networks}
\author{Ashok Ajoy}
\email{ashokaj@mit.edu}
\affiliation{Department of Nuclear Science and Engineering and Research
Laboratory of Electronics,
Massachusetts Institute of Technology, Cambridge, MA, USA}
\author{Paola Cappellaro}
\email{pcappell@mit.edu}
\affiliation{Department of Nuclear Science and Engineering and Research
Laboratory of Electronics,
Massachusetts Institute of Technology, Cambridge, MA, USA}
\begin{abstract}
Spin chains have been proposed as wires to transport information between
distributed registers in a quantum information processor. 
Unfortunately, the challenges in manufacturing 
linear chains with engineered couplings has hindered experimental implementations. 
Here we present  strategies to achieve perfect quantum information transport in arbitrary spin networks. 
 Our proposal is based on the weak coupling limit for pure state transport, where information is transferred between two end-spins that are
 only weakly coupled to the rest of the network.
This regime allows disregarding the complex, internal dynamics of the bulk network and relying on virtual transitions or on the coupling to a single bulk eigenmode. 
We further introduce control methods capable of tuning the transport process and achieve perfect fidelity with limited resources, involving only manipulation of the end-qubits.
These strategies could be thus applied not only to engineered systems with relaxed fabrication precision, but also to naturally occurring networks; specifically, we discuss the practical implementation of quantum state transfer between two
separated nitrogen vacancy (NV) centers through a network of nitrogen
substitutional impurities.
\end{abstract}
\maketitle

%==================================================
Transport of quantum information between distant qubits is an essential task for quantum communication~\cite{Kimble08} and quantum computation~\cite{Cirac97}. 
Linear spin chains have been  proposed~\cite{Bose03}  as quantum wires to connect  distant computational units of a distributed quantum processor. 
This architecture would  overcome the  lack of local addressability of naturally occurring spin networks by separating in space the computational  qubit registers while relying on free evolution of the spin wires to transmit information among them. 
Engineering the coupling between  spins  can improve the transport fidelity~\cite{Christandl04}, even allowing for \textit{perfect} quantum state transport (QST), but it is difficult to achieve in experimental systems.  
Remarkable work~\cite{Wojcik05,Li05} found relaxed coupling engineering requirements -- however, even these proposals  still required  linear chains with nearest-neighbor couplings~\cite{Gualdi08,Yao11} or  networks will all equal couplings~\cite{Wojcik07}. 
These requirements  remain too restrictive to allow an experimental
implementation, since manufacturing highly regular networks is challenging with
current technology~\cite{Strauch08,Hirjibehedin06,Spinicelli11,Ajoy12b}.

Here we describe strategies for achieving QST  between separated ``end''-spins in an \textit{arbitrary} network topology.
We employ the weak-coupling regime \cite{Wojcik05, Li05,Gualdi08,Yao11,Wojcik07,Banchi11},  where the end-spins are engineered to be weakly coupled to the bulk of the network. We  describe the transport dynamics via a perturbative approach, identifying two different regimes. 
Perfect transport can  be achieved by setting the end-spins far off-resonance from the rest of the network -- transport  is then driven by a second-order
 process and hence is slow, but it  requires no active control. 
Faster transport is reached by bringing  the end spins in resonance with a  mode of the bulk of the network, effectively creating a $\Lambda$-type network~\cite{Ajoy12}, whose dynamics we characterize completely. We further introduce a simple control sequence that ensures perfect QST by properly \textit{balancing} the coupling of the end-spins to the common bulk mode, thus allowing perfect and fast state transfer. 
Finally, we investigate the scaling of QST in various types of networks and  discuss practical implementations  for QST between separated nitrogen vacancy (NV) centers in diamond~\cite{Dutt07,Jelezko06} via randomly positioned electronic Nitrogen impurities~\cite{Barklie81,Hanson08,Cappellaro11,Yao12}.

%%%%%%%%%%%%%%%%%%%%%%%%%%%%%%%%%%%%%%%%%%%%%%%%%%%%%%%%%%%%%%%%%%%%%%%%%%
\begin{figure}
 \centering
\includegraphics[scale=0.25]{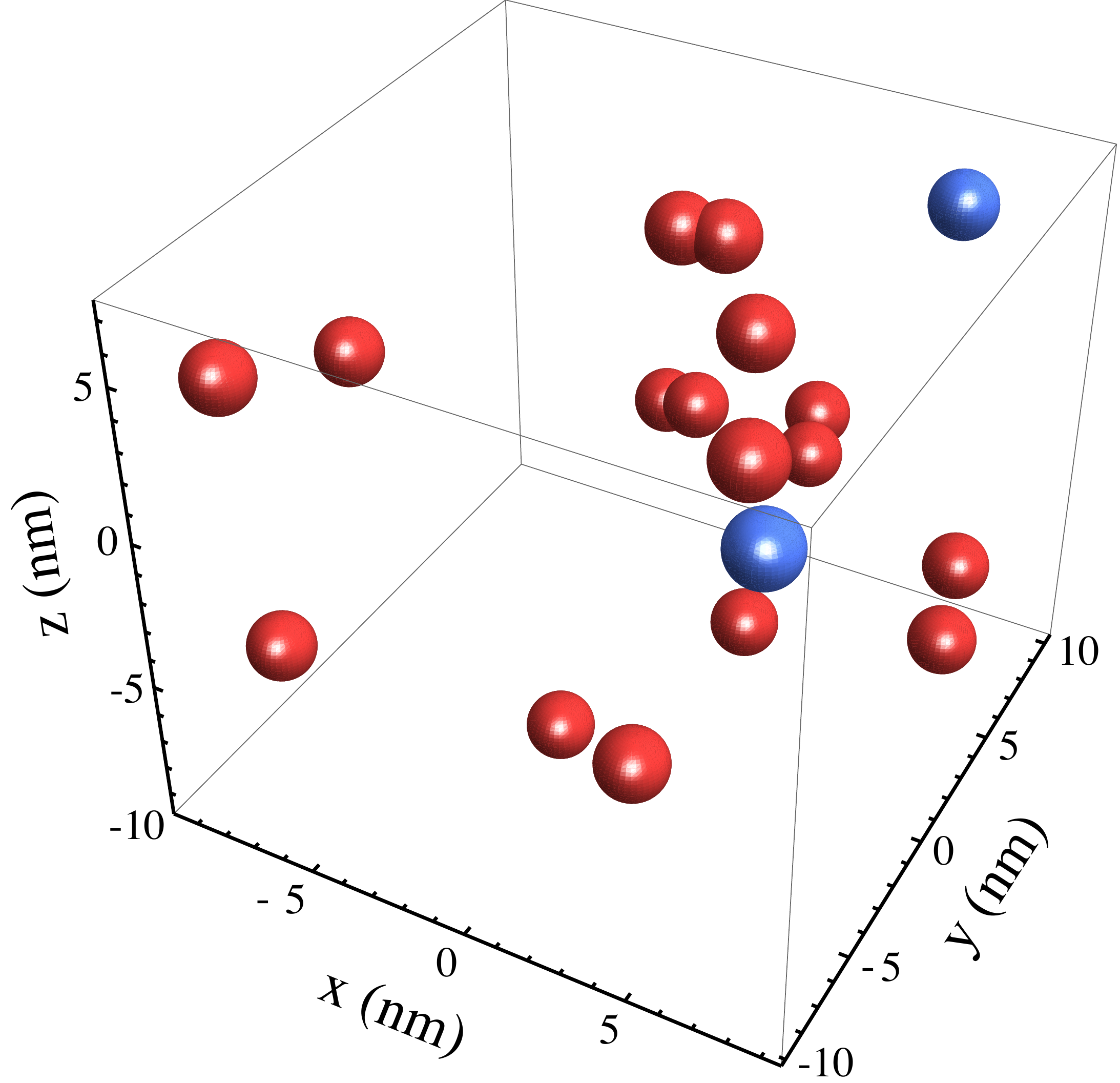}
\caption{Example of spin network, consisting of NV centers (blue spheres) and
P1 centers (red) in a diamond lattice. The network is given by randomly positioned 
 P1 centers in the diamond lattice with a concentration of $0.2$ ppm and a $5\%$ conversion
efficiency to NV. The proposed strategies enable perfect quantum state transfer between the two NV spins in this naturally occurring topology
of P1 centers~\cite{Cappellaro11,Yao12}.}
\label{fig:network}
\end{figure}

%%%%%%%%%%%%%%%%%%%%%%%%%%%%%%%%%%%%%%%%%%%%%%%%%%%%%%%%%%%%%%%%%%%%%%%%%%

\paragraph*{Spin Network --} The system (\zfr{network}) is an $N$-spin network, whose
nodes represent spins-$\half$ and whose edges $H_{ij}$ are the 
Hamiltonian coupling spins $i$ and $j$. We consider the isotropic XY Hamiltonian,  $H_{ij}= (S_i^{+}S_j^{-} +
S_i^{-}S_j^{+})$, with $S_j^{\pm}=\frac{1}{2}(S^x_j \pm iS^y_j)$, which has been widely studied for quantum transport
\cite{Ajoy12,Cappellaro07l,Bose03, Benjamin03}.
We further assume that two nodes, labeled 1 and $N$, can be partially controlled and read out, independently from the \textit{bulk} of the network: we will consider QST between these \textit{end} spins. 
For perfect transport, an excitation created at the location of spin $1$ should be transmitted without distortion to the position of  spin $N$ upon evolution under $H$.  We characterize the efficiency of transport by the fidelity, 
$F(t)=|\bra{N}e^{-iHt}\ket{1}|^2$, where $\ket j$ represents a single excitation $\ket{1}$ at spin $j$, while all other spins are in the ground state $\ket{0}$. 

\paragraph*{Weak-coupling regime --} While optimal fidelity has been obtained
for particular, engineered networks (mainly 1D, nearest-neighbor chains), here
we consider a completely arbitrary bulk network, $H_B$. To ensure perfect
transport, we  work in the weak-coupling regime for the end-spin coupling $H_e$.
By engineering  appropriate  weights $\epsilon, \boldsymbol\beta$, we thus impose $\epsilon\|H_e\|\ll\boldsymbol\beta\|H_B\|$, where 
 $\|\cdot\|$ is a suitable matrix norm~\cite{Bhatia96b}.

%%%%%%%%%%%%%%%%%%%%%%%%%%%%%%%%%%%%%%%%%%%%%%%%%%%%%%%%%%%%%%%%%%%%%%%%%
\begin{figure}
 \centering
\includegraphics[scale=1.3]{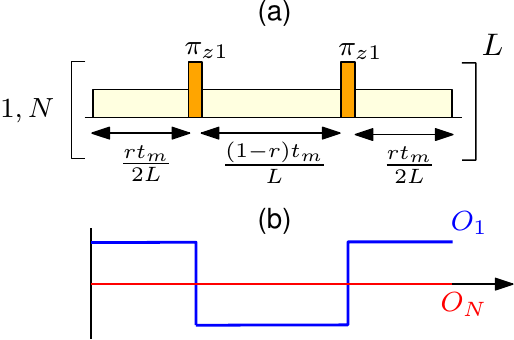}
\caption{(a) On-resonance balancing sequence applied to the end spins for
perfect 
quantum transport. 
Yellow blocks indicate  microwave irradiation that brings the end spins on resonance with the bulk network.
The resulting $\Lambda$-network is in general unbalanced, but  appropriately
placed $\pi$-pulses (orange) on the end-spin with higher mode overlap (here spin
1)  can balance the overlaps $O_{(1,N)}$~\cite{ajoy2011}. 
(b) The $\pi$-pulses invert the sign of  spin-1 coupling to the bulk mode in the
toggling frame such that  $O_{1,N}$ become equal on average. 
For the network of \zfr{network}, $|O_1|\!>\!|O_N|$ and $r\!=\!1/2(1+|O_N/O_1|) = 0.5501$. 
The  sequence is symmetrized \cite{Levitt08,Suzuki90} and repeated for $L$
cycles. }
\label{fig:seq}
\end{figure}
%%%%%%%%%%%%%%%%%%%%%%%%%%%%%%%%%%%%%%%%%%%%%%%%%%%%%%%%%%%%%%%%%%%%%%%%%%
Intuitively, we expect  the weak coupling regime to achieve perfect transfer since it imposes two rates to the spin dynamics:
the bulk spins evolve on a ``fast'' time scale while the end-spin dynamics is 
``slow''. The end spins inject information into the bulk, which evolves so
quickly that information spreads everywhere at a rate much faster
than new information is fed in, allowing an adiabatic elimination of the
information quantum walk in the bulk network~\cite{Ajoy12,Cohen-Tannoudji92b}. 
Although high fidelity can be reached, this \textit{off-resonance} transport is very slow. 

A different strategy, and a faster rate,  for information transport  is achieved by bringing  the end-spins  \textit{on resonance} with an
eigenmode of the bulk -- the weak coupling ensuring greater overlap with a
single (possibly degenerate) mode. 
The system reduces to a $\Lambda$-network~\cite{Ajoy12}, where coupled-mode theory ensures perfect transport
if both ends have equal overlaps with the bulk mode~\cite{Haus91, Synder83}, a condition that we will show can be engineered by  the control sequence in \zfr{seq}.

To make more rigorous our  intuition of the weak regime, we describe transport  via a perturbative treatment.  
For convenience we consider normalized matrices, $\|H_B\|\!=\!\|H_e\|\!=\!1$, and introduce the network adjacency matrix, $A\!=\!\boldsymbol\beta A^B\!+\!\epsilon A^e$, 
which  describes the coupling networks of the system Hamiltonian $H\!=\!\boldsymbol\beta H_B + \epsilon H_e$.
Transport in the single excitation subspace is fully described by  $A$~\cite{Christandl04}, thus the
fidelity can be written as $F(t)=|\bra{N}e^{-iAt}\ket{1}|^2$, where the vectors $\ket{j}$ now 
represent the node basis in the $N\times N$ network space. 
We use a Schrieffer-Wolff  transformation~\cite{Schrieffer66,Bravyi11} and
its truncation to first order in $\epsilon/\boldsymbol\beta$ to define an effective
adjacency matrix, $A'=e^SAe^{-S}\approx A+\half[S,A]$, which drives the evolution.
Setting $S$ so that $[A^B,S]=\frac{\epsilon}{\boldsymbol\beta}A^e$, we have
$A'\approx\boldsymbol\beta A^B+\frac\epsilon2A^S$, where $A^S\!=\![S,A^e]$ can be evaluated explicitly. 

\paragraph*{Off-resonance QST --}
Consider first the case where the eigenvalues $\{E_j^B\}$ of $A^B$ are non-degenerate,
except for $E_1^B\!=\!E_N^B\!=\!0$ (associated with the end-spin subspace). 
 We can  fix the energy eigenbasis $\{\ket{v_k}\}$ of $A^B$  by setting $\ket{v_1}=\ket{1}$ and
$\ket{v_N}=\ket{N}$. In  this basis  the  structure of the matrix $A^e$ is  preserved, non-zero terms connecting only the ends to the bulk,
$A^e_{\ell,j}=0$ for $\ell\neq1,N$. A general element of $A^S$ can be written as,
%\begin{equation} 
\[A^S_{ij}=\frac{\epsilon^2}{\boldsymbol\beta}\sum_k
A^e_{ik}A^e_{kj}\left(\frac1{E_i^B-E_k^B}+\frac1{E_j^B-E_k^B}\right)\]
%\label{eqn:AS}
%\end{equation} 
Given the form of $A^e$, we have, for $\{i,j\}\neq1,N$,
\begin{equation} 
A^S_{ij}=\frac{\epsilon^2}{\boldsymbol\beta}
(A_{i1}A_{1j}+A_{iN}A_{Nj})\left(\frac1{E_i^B}+\frac1{E_j^B}\right).
\label{eqn:ASbulk}
\end{equation} 
Also, setting $k\neq1,N$ and $\{\zeta,\xi\}\in\{1,N\}$ we have 
\begin{equation}  
A^S_{\zeta\xi}=-\textstyle\frac{2\epsilon^2}{\boldsymbol\beta}\sum_k A^e_{\zeta k}A^e_{k \xi}/{E^B_k},
\label{eqn:ASend}
\end{equation} 
while if $A^e_{\xi,\zeta}=0$ elements between the end and bulk are zero, $A^S_{\zeta k}=0$.
%, since then we can simplify \zr{AS} to
%\[\begin{array}{lcr}
%A^S_{\zeta k}&=&
%\frac{\epsilon^2}{\boldsymbol\beta}A^e_{\zeta\zeta}A^e_{\zeta k}\left(\frac1{-E_\zeta^B}+\frac1{
%E_k^B-E_\zeta^B}\right)\\
%&+&\frac{\epsilon^2}{\boldsymbol\beta}
%A^e_{\zeta\xi}A^e_{\xi k}\left(\frac1{-E_\xi^B}+\frac1{E_k^B-E_\xi^B}\right).
%\end{array}\]
%%%%%%%%%%%%%%%%%%%%%%%%%%%%%%%%%%%%%%%%%%%%%%%%%%%%%%%%%%%%%%%%%%%%%%%%%%
\begin{figure}
 \centering
\includegraphics[scale=0.23]{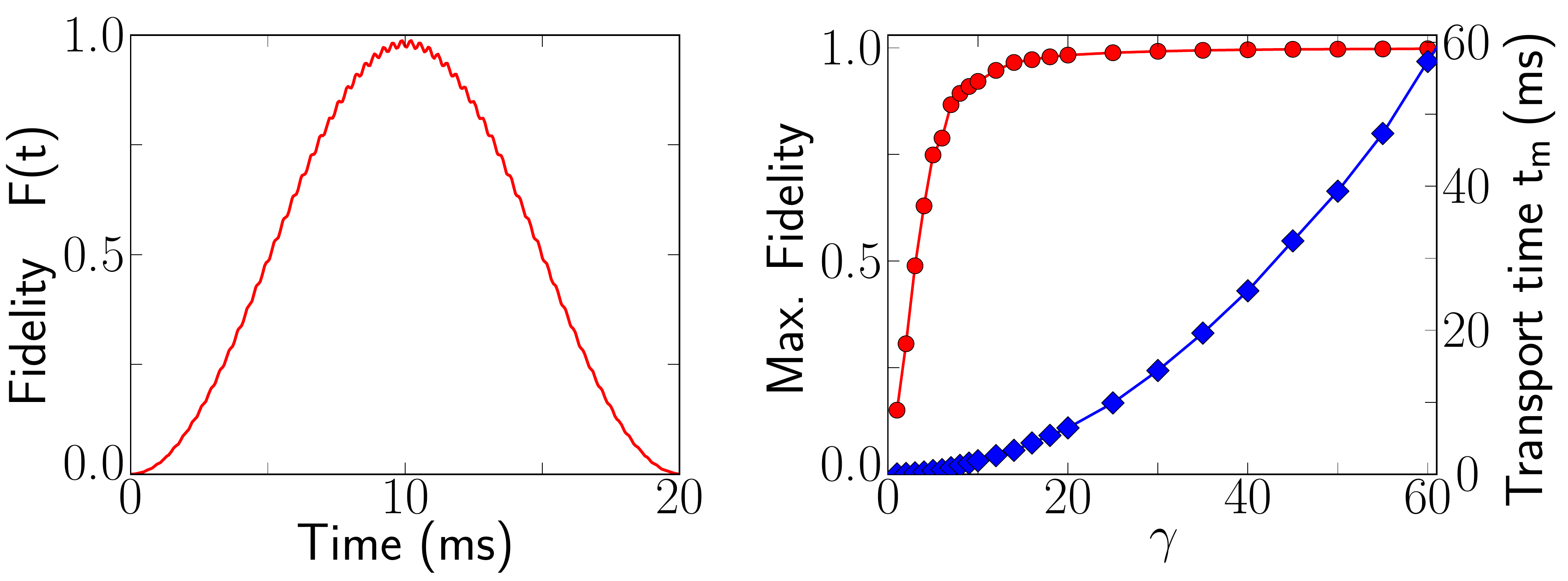}
\caption{(a) Off-resonance transport fidelity for the network in \zfr{network}  for $\gamma=25$. 
Perfect transport occurs, but on a slow time scale,  an order of magnitude longer than the on-resonance balanced case
(\zfr{on-res}). 
(b) Increasing $\gamma$ improves transport fidelity (red circles) but also
increases the time required for perfect transport (blue diamonds).}
\label{fig:off-res}
\end{figure}
%%%%%%%%%%%%%%%%%%%%%%%%%%%%%%%%%%%%%%%%%%%%%%%%%%%%%%%%%%%%%%%%%%%%%%%%%%
 Hence $A^S$ can be partitioned into a term
with support only in the bulk subspace (Eq.~\ref{eqn:ASbulk}) and a second term with support only in the end-spin subspace (Eq.~\ref{eqn:ASend}), while
there is no end-bulk coupling. 
To first order approximation, it is only the latter term that drives QST -- the transport
happens via a direct coupling between the
end-nodes. Since this effective coupling is  mediated by the bulk via virtual
transitions, its rate is proportional to $\epsilon^2/\boldsymbol\beta$. The
fidelity of transport is determined entirely by the effective detuning of the
two end-spins, $\alpha=(A^S_{11}-A^S_{NN})/2$:
\begin{equation}
F(t)=\frac{(A^S_{1N})^2}{(A^S_{1N})^2+\alpha^2}\sin^2\left(t\sqrt{(A^S_{1N}
)^2+\alpha^2}\right)
\label{eqn:fidoffres}
\end{equation}
If we can modify the end-spin Hamiltonian by adding a term
$H_{\text{off}}=-2(\omega_1S^z_1+\omega_NS^z_N)$, such that $A^S_{11}+\omega_1=A^S_{NN}+\omega_N$,
perfect quantum transport is ensured at $t_m=\pi/(2A^S_{1N})$. 
This energy shift could be obtained by locally tuning the magnetic field or by applying local AC driving, ensuring the desired energy in the rotating frame (similar to the Hartman-Hahn scheme~\cite{Hartman62}).
Transport fidelity also depends on the goodness of the first order approximation, increasing with $\boldsymbol\beta/\epsilon$ as shown in \zfr{off-res}(b) at the cost of longer transport times.

\paragraph*{On-resonance QST} -- Transport can be made faster if the end spins are on
resonance with one non-degenerate mode of the bulk $\ket{v_d}$ (we will consider the degenerate case in~\cite{SOM}).
Resonance happens if the corresponding eigenvalue $E_d^B=0$ 
or it can be enforced  by adding an energy shift to the end spins  to set $A^e_{11}=A^e_{NN}=\boldsymbol\beta/\epsilon E^B_d$. 
Transport then occurs at a rate proportional to $\epsilon$, as driven by the adjacency matrix
$A^d$,  the projection of $A^e$ in the degenerate subspace:
\[
A^d=\bra{1}A^e\ket{v_j}\ket{1}\!\!\bra{v_d}+\bra{N}A^e\ket{v_d}\ket{N}\!\!\bra{v_d}+h.c.
\]
We note that the goodness of this approximation depends on the gap between the selected resonance mode and the other bulk modes.
In the node basis, $A^d$  forms an effective \textit{$\Lambda$-network}~\cite{Ajoy12}, coupling the end-spins
 with each spin of the bulk:
\begin{equation}
A^d=\sum_j (\delta_{1j} \ket{1}\!\!\bra{j}+\delta_{Nj}\ket{N}\!\!\bra{j}+h.c.),
\label{eqn:Ad}
\end{equation}
where $\delta_{(1,N)j}=\braket{v_d}j\bra{(1,N)}A^e\ket{v_d}$.  Note that importantly we have $\delta_{1j}/\delta_{Nj}=\text{cst.}$, $\forall j$. 
Transport in such $\Lambda$-networks  occurs at only four frequencies~\cite{SOM},  $F(t)=w_0 +\sum_{m=1}^{4} w_m\cos(f_mt)$ with,
\begin{equation} 
f_{1,2} = 2\sqrt{S^2 \mp \sqrt{S^4 - \Delta^4}}\:;\:f_{3,4} = \sqrt{2(S^2 \mp \Delta^2)},
\end{equation} 
where,
\begin{eqnarray} 
S^2 &=&  \sum_j\half(\delta_{1j}^2 + \delta_{jN}^2)\nonumber\\
\Delta^4 &=& \sum_{j<k}(\delta_{1j}\delta_{kN} -\delta_{jN}\delta_{1k})^2 \\
\delta^2 &=& \sum_j \delta_{1j}\delta_{jN}\nonumber
\label{eqn:Sdefn}
\end{eqnarray} 
Physically, $S\sim\|A^d\|$ sets the energy scale of the resonant mode, while
$\Delta$ quantifies the {\it relative} detuning between  different
$\Lambda$-paths~\cite{Ajoy12,SOM}. The coefficients $w_i$ are found to be 
$w_0=-w_3=-w_4=\frac{\delta^4}{4(S^4-\Delta^4)}$, $w_1=w_2={w_0}/2$, giving the
analytical expression for QST in a $\Lambda$-network,
\begin{equation} 
F(t)\!=\!\frac{\delta^4}{S^4-\Delta^4}\sin^2\!\left(\!\!\sqrt{\frac{S^2 +\Delta^2}2}t\!\right)\!\sin^2\!\left(\!\!\sqrt{\frac{S^2 -\Delta^2}2}t\!\right)
\end{equation} 
Perfect QST  requires $\Delta=0$ and $\delta=S$. The first condition entails $\delta_{1j}/\delta_{jN}=\textrm{cst.}$  $\forall j$, which is always satisfied by $A^d$ if the resonant bulk mode is {non-degenerate}. 
On the other hand, $\delta=S$ requires a \textit{balanced} network,  $\delta_{1j}=\delta_{jN}$ $\forall j$. 
For the reduced adjacency matrix $A^d$ this condition is satisfied when both end-spins have equal overlap with the resonant eigenmode, $\bra{1}A^e\ket{v_d}=\bra{N}A^e\ket{v_d}$, and we show below that this can  be always arranged for a non-degenerate mode by a simple control sequence (see \zfr{seq}). 
In the balanced case the fidelity  simplifies to $F(t)=\sin(St/\sqrt2)^4$, which leads to perfect QST at $t_m=\pi/(\sqrt2S)$, as if it were a 3-spin chain~\cite{Christandl04}.

%%%%%%%%%%%%%%%%%%%%%%%%%%%%%%%%%%%%%%%%%%%%%%%%%%%%%%%%%%%%%%%%%%%%%%%%%%
\begin{figure}[t]
 \centering
\includegraphics[scale=0.23]{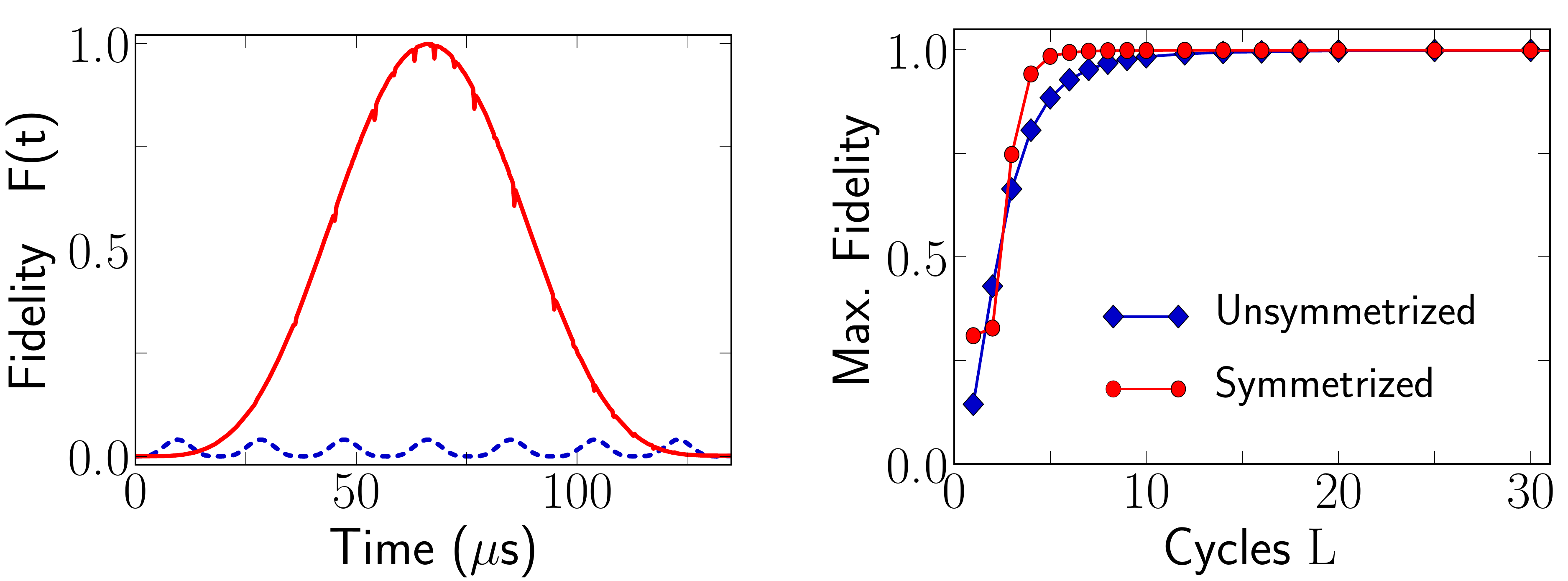}
\caption{(a) On-resonance transport fidelity for the network of \zfr{network}, 
with (red) and without (blue, dotted) balancing,  with $\gamma=1$ and $L=20$.
Almost perfect transport occurs in the balanced case, obtained by the control
sequence in \zfr{seq}.
(b) Increasing the number of cycles $L$ improves the Trotter approximation yielding  enhanced fidelity. 
The symmetrized sequence performs better than the sequence without
symmetrization.
}
\label{fig:on-res}
\end{figure}
%%%%%%%%%%%%%%%%%%%%%%%%%%%%%%%%%%%%%%%%%%%%%%%%%%%%%%%%%%%%%%%%%%%%%%%%%%

\paragraph*{Perfect QST by on-resonance balancing} --
 Unfortunately the  overlaps of the two end spins with the on-resonance  mode, $O_{(1,N)}=\sand{(1,N)}{A^e}{v_d}$, are in general unequal, 
 and the $\Lambda$-network unbalanced. Still, it is possible to \textit{balance} the network by a simple control sequence (\zfr{seq}). 
 Assume for example $O_1>O_N$;   we partition $A^d$ into effective adjacency matrices with couplings only to spins 1 and $N$, 
 $A^d=A^d_1 + A^d_N$.  
 A rotation $e^{-iS^z_1\pi}$ on spin 1 produces $\tilde A^{d}=-A^d_1 + A^d_N$. Then the evolution,
$e^{-iA^drt}e^{-i\tilde A^{d}(1-r)t}\!\approx\!e^{-iA^d_bt}$, with $A^d_b=A^{d}r+\tilde A^{d}(1\!-\!r)$, is balanced on \textit{average} during the period
$t$ if $r=\frac{1}{2}(1+ \labs O_N/O_1\rabs)$. 
The approximation 
improves if one uses $L$ cycles of the control sequence with shorter time intervals (as in a Trotter expansion~\cite{Trotter59,Suzuki90}) and appropriately symmetrizes it  (see \zfr{seq}) to achieve an error ${\cal O}({r^2(1-r)t_m^3}/{L^3})$.
\zfr{on-res}(a)
shows the effect of enhanced, almost perfect, fidelity upon balancing the network of \zfr{network}. 

\paragraph*{Transport time and control requirements} -- 
We now consider the scaling of the weak coupling parameter $\gamma=\frac{\boldsymbol\beta }{\epsilon }$ required for the validity of the perturbative approximation. 
We can fix $\boldsymbol\beta=\|H_B\|$ and consider normalized matrices. Then $\boldsymbol\beta$  scales as $\sqrt{(N-2)(N-3)/3}$ for a random network where all the couplings are uniformly distributed. For the more realistic case where the coupling strength decreases with distance, the scaling is less favorable, e.g.  for a random dipolar coupled network $\boldsymbol\beta$ scales as $\sqrt{2/3(N-3)}/d^3$, where $d$ is the
average lattice constant~\cite{SOM}. 
Similar scaling occurs for regular spin networks, for example those corresponding to crystal lattices \cite{SOM}. 
In general the ratio $\epsilon/\boldsymbol\beta$  decreases with the size of the network (usually as ${\cal O}(\sqrt{N})$), averting the need to reduce the end-couplings by engineering $\epsilon $. 
This is evident in \zfr{on-res}, where $\epsilon =1$ is sufficient to drive perfect quantum transfer. 

In the case of on-resonance balancing  the time at which perfect QST is  achieved is $t_m=\gamma\pi/(\sqrt{2}S)$, where $S=\textrm{min}\{O_1,O_N\}$,
scales linearly with $\gamma$. 
The time is shorter the more symmetrical the end-spins are with respect to the resonant mode \cite{SOM}, since then $|O_1|\approx|O_N|$.
For the off-resonance case the time is $t_m=(\pi\boldsymbol\beta /2\epsilon ^2)[E_\ell^B/(A^e_{1\ell}A^e_{k\ell})]$, where
$E^B_\ell=\textrm{min}\{\labs E^B_k\rabs\}$ is the minimum eigenvalue of the bulk.
In general this second order transport process takes an order of magnitude longer time than the on-resonance case (\zfr{off-res}).

%%%%%%%%%%%%%%%%%%%%%%%%%%%%%%%%%%%%%%%%%%%%%%%%%%%%%%%%%%%%%%%%%%%%%%%%%%
\begin{figure}
 \centering
\includegraphics[scale=0.25]{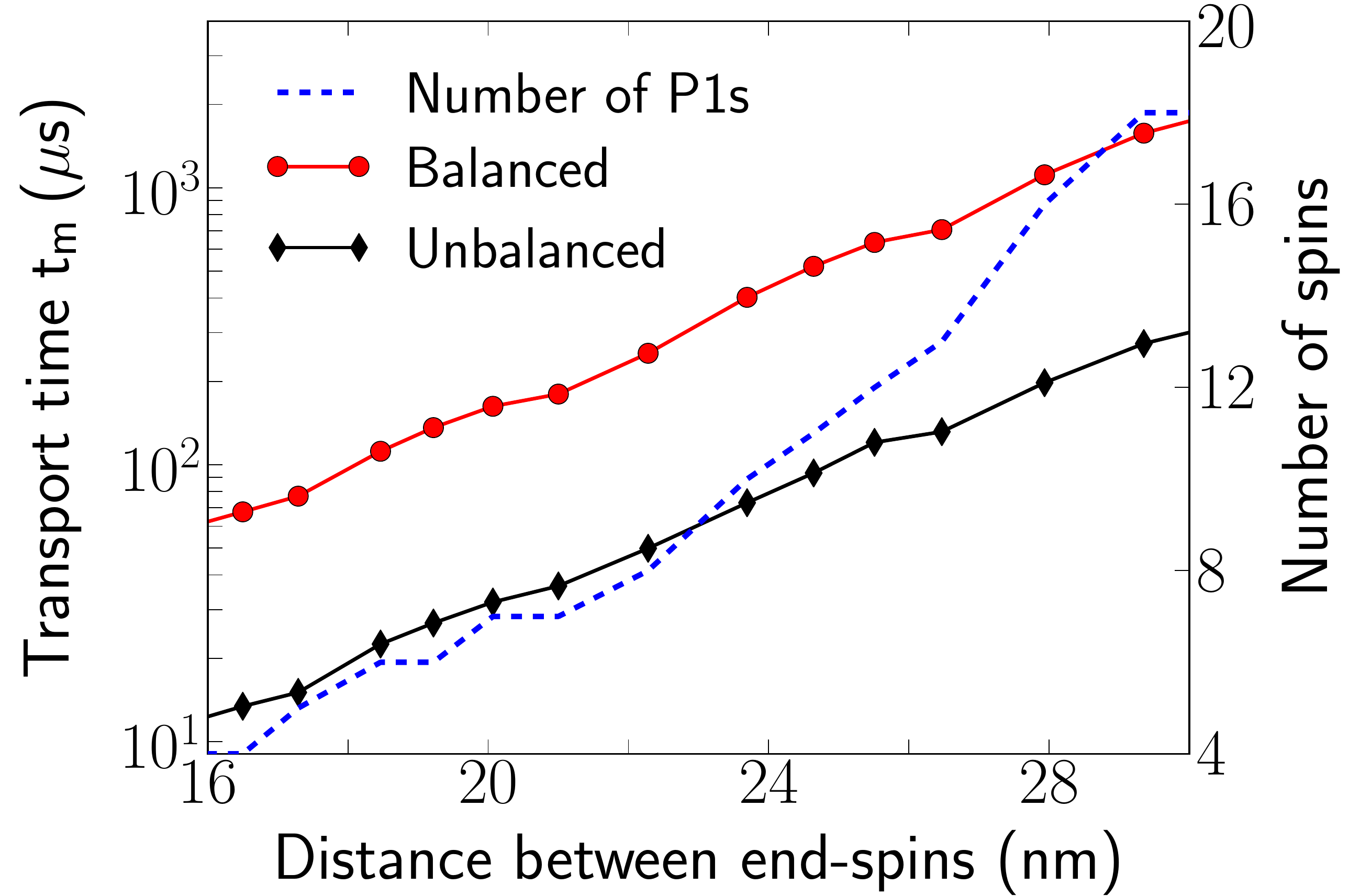}
\caption{Time for optimal transport  in dipolar coupled
spin networks of P1 centers in diamond. The optimal
time was calculated from the average over 5000 random lattice realizations of
P1 centers of density 10ppm; we considered transport between two NV centers
located at increasing distance. The red line (circles) is the time
required to achieve a transport fidelity of at least 99\% via resonant
balancing. The black line (diamonds) is the time at which optimal fidelity is
achieved for on-resonant transport without balancing. The 
fidelity is however quite low ($\sim\!15\%$) in this case. The blue dashed line
shows the number of P1 centers in the network, $N\!-\!2$.}
\label{fig:time}
\end{figure}

%%%%%%%%%%%%%%%%%%%%%%%%%%%%%%%%%%%%%%%%%%%%%%%%%%%%%%%%%%%%%%%%%%%%%%%%%%

Finally, let us estimate the  resources needed to impose the  end-spin energy shifts as required for perfect QST, for example by a continuous  irradiation during $t_m$ (\zfr{seq}).
In the on-resonance case, the end-spin energy should be set equal to a bulk mode, $E^B_d$. 
Selecting the highest bulk eigenmode, which has usually the largest gap to the other modes, $E^B_d$ scales as ${\cal O}(N)$ for random networks, but it is  constant,  $\sim1.6/d^3$, for dipolar coupled networks \cite{SOM}. 
Off-resonance transport requires instead a shift of $A^S_{11} - A^S_{NN}$, where $A^S_{11,NN} = \braket{A^e_{1,N}}{\ell}^2/\gamma^2E^B_{\ell}$, where $E^B_{\ell}$ is the minimum bulk eigenmode. 
Hence the control required in this case is about an order of magnitude lower than in the resonant case.

An experimentally important quantum computing architecture is that of nitrogen
vacancy (NV) centers in diamond. \zfr{time} shows the scaling of the transport
time between two separated NV centers via a bulk network consisting of randomly
positioned nitrogen impurities (P1 centers).  Some experimental challenges remain:  the limitation of pure state
transfer requires that the system is at low temperature or that P1 centers are indirectly polarized by the NV centers; the
requirement of using the isotropic XY Hamiltonian requires additional
control for it to be generated from the natural dipolar Hamiltonian (for
example via a combination of gradient and TOCSY mixing~\cite{tocsy,Yao12,ajoy-unpub}).
Still, considering that dephasing times in excess of 100$\mu$s are routinely
achievable~\cite{Stanwix10,Balasubramanian09}, the balanced transport
scheme may be experimentally viable for quantum communication in these
architectures. 

\paragraph*{Conclusion -- }In this paper, we showed that perfect quantum state transfer
can be engineered even in the case of arbitrarily complicated network
topologies, if the ends of the network are
weakly coupled to the bulk. Transport speed can be improved by bringing the end
spins on resonance with a common mode of the bulk network. Alternatively, it is
possible to achieve unit transport fidelity, with lower control requirements,
but on a longer time scale, by detuning the end-spins
off-resonance. These transport strategies may allow the interlinking of quantum
registers in a quantum information processor with very relaxed fabrication requirements.
\paragraph*{Acknowledgments -- }
This work was partially funded by NSF under grant DMG-1005926 and by AFOSR YIP.

\bibliographystyle{apsrev}
\bibliography{Biblio-p-url,Biblio,mixed}

\onecolumngrid
\pagebreak
\appendix
\section{Transport for end-spins on-resonance with a degenerate mode}
In the main text we described transport when the end spins are on resonance with
an eigenmode of the bulk with eigenvalue $E^B_d$.  Here we generalize to the
situation where the eigenvalue is degenerate, with eigenvectors
$\ket{\alpha_m}$.  
The projector onto the degenerate eigenspace is then,
\begin{equation}
P = (\ketbra{1}{1}+ \ketbra{N}{N}) + \sum_{m\in M}\ketbra{\alpha_m}{\alpha_m}
\equiv (\ketbra{1}{1}+ \ketbra{N}{N}) + P_M ,
\end{equation}
and, to first order, transport is driven by the projection of the adjacency
matrix $A^e$ into the this subspace, i.e., $A^d=P^{\dag}A^eP$.
Now we consider that $A^e$ and $A^B$ have the following forms, where
``$\times$''
denotes a non-zero element:
\begin{equation}
A^e = \left[
\begin{array}{cccccc}
{}&{\times}&{\times}&{\times}&{\times}&{}\\
{\times}&{}&{}&{}&{}&{\times}\\
{\times}&{}&{}&{}&{}&{\times}\\
{\times}&{}&{}&{}&{}&{\times}\\
{\times}&{}&{}&{}&{}&{\times}\\
{}&{\times}&{\times}&{\times}&{\times}&{}
\end{array}
\right]
\:;\:\qquad
A^B = \left[
\begin{array}{cccccc}
{\ }&{}&{}&{}&{}&{}\\
{}&{}&{\times}&{\times}&{\times}&{\ }\\
{}&{\times}&{}&{\times}&{\times}&{}\\
{}&{\times}&{\times}&{}&{\times}&{}\\
{}&{\times}&{\times}&{\times}&{}&{}\\
{}&{}&{}&{}&{}&{}
\end{array}
\right].
\end{equation}
\renewcommand{\arraystretch}{2}
It follows  that
\begin{equation}
\sand{1}{A^e}{1} = 0\:;\:\sand{N}{A^e}{N} =
0\:;\:\sand{\alpha_{m'}}{A^e}{\alpha_m} = 0,
\end{equation}
yielding 
\begin{equation}
A^d=\sum_m\sand{1}{A^e}{\alpha_m}(\ketbra{1}{\alpha_m} + \ketbra{\alpha_m}{1}) +
\sum_m
\sand{N}{A^e}{\alpha_m}(\ketbra{N}{\alpha_m} + \ketbra{\alpha_m}{N}),
\end{equation}
where we have used the fact that $A^e$ is real, and hence
$\sand{1}{A^e}{\alpha_m}
= \sand{\alpha_m}{A^e}{1}$. Let us define the end connection vectors,
\begin{equation}
A^e\ket{1} = \ket{n_1}\:;\:A^e\ket{N} = \ket{n_N}.
\end{equation}
Then,
\begin{eqnarray}
A^d&=&\sum_m\braket{n_1}{\alpha_m}(\ketbra{1}{\alpha_m} + \ketbra{\alpha_m}{1})
+
\sum_m\braket{n_N}{\alpha_m}(\ketbra{N}{\alpha_m} +
\ketbra{\alpha_m}{N})\nonumber\\
&=&\sum_j\sum_m\braket{n_1}{\alpha_m}\left(\braket{\alpha_m}{j}\ketbra{1}{j} +
\braket{j}{\alpha_m}\ketbra{j}{1}\right) + 
\sum_j\sum_m\braket{n_N}{\alpha_m}\left(\braket{\alpha_m}{j}\ketbra{N}{j} +
\braket{j}{\alpha_m}\ketbra{j}{N}\right)\nonumber
\end{eqnarray}
or simplifying,
\begin{equation}
A^d=\sum_j (\delta_{1j} \ketbra{1}{j}+  \delta_{jN}\ketbra{N}{j}+h.c.),
\end{equation}
where $\delta_{(1,N)j} = \sand{n_{(1,N)}}{P_M}{j}$ 
is the overlap of the end vector $n_1$ and the node $j$ in the resonant
subspace.

To achieve  balanced on-resonance transport we require that
$\delta_{1j}=\delta_{jN}$ for
all $j$, which implies that both the end-vectors have equal projections in the
resonant subspace,
\begin{equation}
P_M\ket{n_1} = P_M\ket{n_N}.
\end{equation}

\section{Transport Fidelity for $\Lambda$-networks}

%%%%%%%%%%%%%%%%%%%%%%%%%%%%%%%%%%%%%%%%%%%%%%%%%%%%%%%%%%%%%%%%%%%%%%%%%%%%%%%
\begin{figure}
 \centering
\includegraphics[scale=0.8]{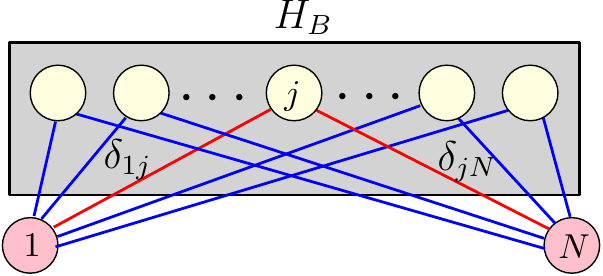}
\caption{(Color online) A general detuned $\Lambda$-network with multiple
$\Lambda$-paths between the end spins. Each leg of any of these $\Lambda$-paths
may
have an arbitrary detuning. For the representative case of the $\Lambda$-path
going through
node $j$ (red), these detunings are $\delta_{1j}$ and $\delta_{jN}$.
}
\label{fig:13}
\end{figure}
%%%%%%%%%%%%%%%%%%%%%%%%%%%%%%%%%%%%%%%%%%%%%%%%%%%%%%%%%%%%%%%%%%%%%%%%%%%%%%%

Here we derive  the maximum transport fidelity for a
$\Lambda$-type network. $\Lambda$-type networks are interesting because the
effective Hamiltonian of more complex networks 
reduces to $\Lambda$-network Hamiltonian in the weak-coupling regime (as shown
in the previous section),  but we will consider here the general case. 

For any network of adjacency matrix $A$,  the fidelity function
$F(t)\!=\!|\!\bra{N}e^{-iAt}\ket{1}\!|^2$ has a simple expression in  the
eigenbasis of $A$,
\begin{equation}
F(t)\!=\sum_{k,\ell}\braket{\ell}{N}\braket N{k}\braket{k}1\braket
1\ell\cos\left(E_{k} -
E_\ell\right)t.
\label{eqn:A4.3}
\end{equation}
This shows that the fidelity can be written as the sum 
\begin{equation}
F(t) =\sum_{n}w_n\cos(f_nt),\qquad\textrm{with}\quad f_n = E_k - E_{\ell},
\label{eqn:A4.4}
\end{equation}
that is, the  frequencies  are  differences between eigenvalues of $A$, for
which  the corresponding eigenvectors $\ket{\ell}$ and $\ket{k}$  have non-zero
overlap with $\ket{1}, \ket N$.

We now consider a  general $\Lambda$-network with  multiple $\Lambda$-paths that
connect the end-spins (see \zfr{13}) and we will restrict the analysis to the
adjacency matrix obtained in the {\it on-resonance} case only later. 
We write the adjacency matrix in terms of the  coupling strength  $\delta_{1j}$
and $\delta_{jN}$ between the end spin and each $j^{th}$ spin in the bulk, which
form the $\Lambda$ path: 
\begin{equation}
A^d=\sum_j\Lambda_j\:,\qquad
\Lambda_j = \delta_{1j}\left(\ket{j}\!\!\bra{1}+\ket1\!\!\bra j\right)+
\delta_{jN}(\ket{j}\!\!\bra{N}+\ket N\!\!\bra j).
\label{eqn:A5.1}
\end{equation}

Our strategy for finding the transport fidelity in $\Lambda$-networks is to
first determine the eigenvalues of the adjacency matrix in \zr{A5.1} and hence
the
possible frequencies at which  information transport can occur. Then, we will
use a series expansion to find an explicit expression for the fidelity.
\subsection{Frequency of Transport}
The eigenvalues of the  adjacency matrix in \zr{A5.1} are
\begin{eqnarray}
\lambda_0&=&0,\qquad (N-4)\text{ degenerate}\nonumber\\
\lambda_{1,2} &=& \pm\sqrt{  S^2 - \sqrt{S^4 -\Delta^4}},\\
\lambda_{3,4} &=&\pm \sqrt{  S^2 + \sqrt{S^4 -\Delta^4}},\nonumber
\label{eqn:A5.2}
\end{eqnarray}
where we defined,
\begin{eqnarray}
S^2 &=& \sum_j S^2_j = \sum_j\half(\delta_{1j}^2 + \delta_{jN}^2)\nonumber\\
\Delta^4 &=& \sum_{j<k}\Delta_{jk}^4 = \sum_{j<k}(\delta_{1j}\delta_{kN}
-\delta_{jN}\delta_{1k})^2 \\
\delta^2 &=& \sum_j \delta_{1j}\delta_{jN}\nonumber
\label{eqn:A5.3}
\end{eqnarray}
%%%%%%%%%%%%%a%%%%%%%%%%%%%%%%%%%%%%%%%%%%%%%%%%%%%%%%%%%%%%%%%%%%
\begin{figure}[b]
 \centering
\includegraphics[scale=0.3]{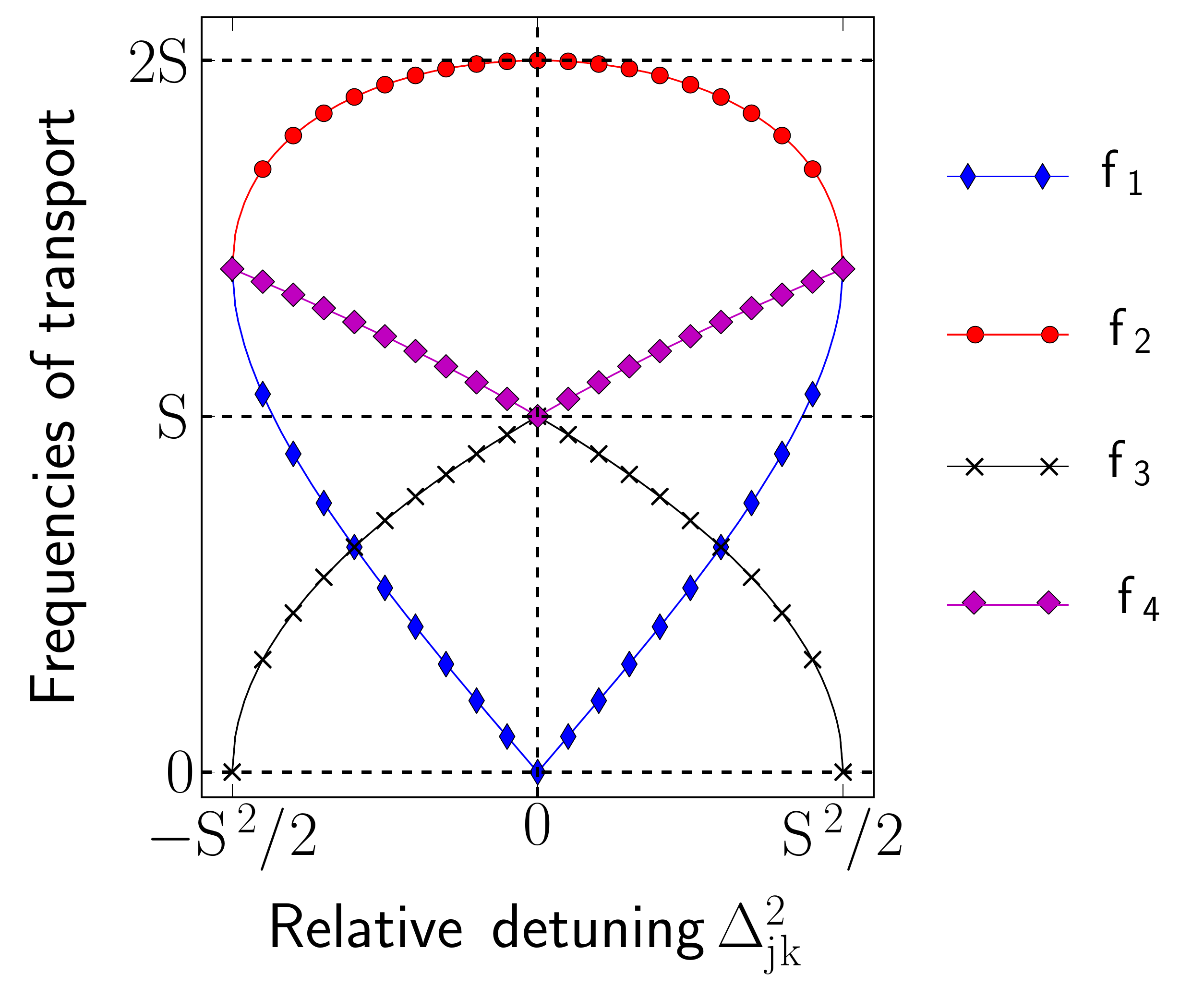}
\caption{(Color online) Positive frequencies of transport fidelity for
a $\Lambda$-network (\zfr{13}) consisting of two $\Lambda$-paths, $1\to j\to N$
and
$1\to k\to N$.
The frequencies are plotted as a function of the relative detuning between the
paths, $\Delta_{jk}^2 = (\delta_{1j}\delta_{kN} - \delta_{jN}\delta_{1k})$. The
actual
transport
also contains symmetric negative frequencies and a DC (zero frequency)
component. In general there are four frequencies of transport, derived in
\zr{A5.5}. Note
that when $\delta_{jN}/\delta_{1j}$ is a constant for both paths, there are only
two
frequencies, $S$ and $2S$, that carry the transport.}
\label{fig:14}
\end{figure}

%%%%%%%%%%%%%a%%%%%%%%%%%%%%%%%%%%%%%%%%%%%%%%%%%%%%%%%%%%%%%%%%%%
While for a general network  the frequencies of transport  are differences
between the eigenvalues of $A^d$, here there are only
four distinct frequencies because of the symmetries in the eigenvalues:
\begin{equation}\begin{array}{l} 
f_0=0,\qquad\qquad f_1=2\lambda_1,\qquad \qquad f_2=2\lambda_3, \\ 
f_3=\lambda_1-\lambda_3=\sqrt{2(S^2-\Delta^2)},\quad
f_4=\lambda_1+\lambda_3=\sqrt{2(S^2+\Delta^2)}.
\end{array}
\label{eqn:A5.5}
\end{equation}

\subsection{Series expansion}
With the frequencies found above, equation \zr{A4.4}  reduces to $F(t)=
\sum_{i=0}^4 w_i \cos{(f_it)}$. To find the parameters $w_i$, we 
equate the Taylor expansion of \zr{A4.4} and  of the fidelity
$F(t)\!=\!|\!\bra{N}e^{-iAt}\ket{1}\!|^2$.  
We only need the first five even power coefficients to fully determine
$\{w_i\}$, giving the series of equations,
\begin{eqnarray}
&&\sum_{i=0}^4w_i\!=\!|\braket {N}{1}|^2=
0\nonumber\\
&&\sum_{i=1}^4w_if_i^2\!=\!-|\sand {N}{A^d}{1}|^2=
0\nonumber\\
&&\frac1{4!}\sum_{i=0}^4w_if_i^4\!=\!\frac14|\sand {N}{\left(
A^{d}\right)^2}{1}|^2=\mathcal C_4
\\
&&\frac1{6!}\sum_{i=0}^4w_if_i^6\!=-\!\frac1{24}\text{Re}[\sand {N}{\left(
A^{d}\right)^2}{1}\sand {N}{\left( A^{d}\right)^4}{1}]=\mathcal C_6
\nonumber\\
&&\frac1{8!}\sum_{i=0}^4w_if_i^8\!=\!\frac1{4!}|\sand {N}{\left(
A^{d}\right)^4}{1}|^2 +\frac1{2\cdot6!}\text{Re}[\sand {N}{\left(
A^{d}\right)^2}{1}\sand {N}{\left( A^{d}\right)^6}{1}]=\mathcal
C_8\nonumber
\label{eqn:weights}
\end{eqnarray}
The expectation values can be evaluated exactly, yielding 
\begin{eqnarray}
{\cal C}_4 &=& \frac14\delta^4\nonumber \\
{\cal C}_6 &=&-\frac1{12}S^2\delta^4\\
{\cal C}_8 &=& \frac1{720}\delta^2\left[ 9S^4 - \Delta^2\right]\nonumber
\label{eqn:A5.7.5}
\end{eqnarray}
For general frequencies $f_i$,  the  coefficients $w_i$ are
\begin{eqnarray}
w_0=-\sum_{j=1}^{4}w_j;&\qquad
w_j = \frac{\mathcal{C}_8 - \sum_{k\neq j}f_k^2\mathcal C_6 +
\sum_{\ell<m;\ell,m\neq
j}f_{\ell}^2f_m^2\mathcal C_4}
{f_j^2\prod_{k\neq j}(f_k^2-f_j^2)},& j>0
\label{eqn:A5.8}
\end{eqnarray}
Using the expressions for the frequencies in \zr{A5.5}, we  find their explicit
expressions in terms of
$S,\Delta$ and $\delta$: 
\begin{equation}
w_0=\frac{\delta^4}{4(S^4-\Delta^4)},\qquad w_1=w_2=\frac{w_0}2,\qquad
w_3=w_4=-w_0
\end{equation}
The fidelity is thus further simplified to 
\begin{equation}
F(t)=\frac{\delta^4}{S^4-\Delta^4}\left[\sin\left(t\sqrt{(S^2
+\Delta^2)/2}\right)\sin\left(t\sqrt{(S^2 -\Delta^2)/2}\right)\right]^2
\end{equation}
\subsection{Fidelity for random and degenerate networks}
Consider the case when the number of nodes is large, and the detunings
$\delta_{1j}$ and $\delta_{jN}$ are sampled from the same distribution, as it
would be in a random network. Then, 
\begin{equation}
\sum_j\delta_{1j}^2 \approx \sum_j\delta_{jN}^2 
\label{eqn:equal-del}
\end{equation}
since the second moments of the random distribution should be equal.
In this situation we have 
\begin{equation}
S^4 - (\Delta^4 + \delta^4)=\frac14\left[\sum_j(\delta_{1j}^2 -
\delta_{jN}^2)\right]^2 = 0
\end{equation}
Then the condition
\begin{equation}
\delta^4 = S^4-\Delta^4 
\label{eqn:perfectfid}
\end{equation}
is satisfied and the fidelity becomes
\begin{equation}
F(t)=\left[\sin\left(t\sqrt{(S^2 +\Delta^2)/2}\right)\sin\left(t\sqrt{(S^2
-\Delta^2)/2}\right)\right]^2
\end{equation}
In the case of resonance to a non-degenerate mode, we have $\Delta=0$ and the
fidelity can reach its maximum $F(t)=\sin({St}/{\sqrt2})^4=1$, for
$t=\pi/\sqrt2S$.

For the case of interest in this work,  a network where the end-spins are on
resonance with a non-degenerate mode, the adjacency matrix of relevance in the
weak regime is the reduced adjacency matrix, $A^d$. As shown above, in this case
we have $\Delta=0$ and the fidelity reduces to
\begin{equation}
F(t)=\left[\frac{\delta}{S}\sin\left(\frac{St}{\sqrt2}\right)\right]^4,
\end{equation}
thus maximum fidelity can  be reached only if the condition \zr{perfectfid} is
satisfied.
For example,  the mirror-symmetric case $\delta_{1j} = \delta_{jN},\ \forall j$
yields the optimal fidelity $F=1$ since then $\delta=S$.

%%%%%%
\section{Estimating matrix norms for different network topologies}
\subsection{Different kinds of networks}
In this section, we  consider different classes of networks, and
estimate the norms of the corresponding adjacency matrices $A$. As described in
the main text, the matrix norm of the bulk adjacency matrix is important in
predicting the transport time. For example, the scaling of transport time is
linear with $\gamma$ in the on-resonance case; and the value of $\gamma$
implicitly depends on the norm of the bulk matrix. Hence a large bulk matrix
causes an intrinsically high $\gamma$, and reduces the control requirements on
the
end-spins. All the networks considered are of $N$ spins, and hence the adjacency
matrices are $N\times N$ matrices.
\begin{enumerate}
\item \textbf{Random network:} The matrix $A$ has random entries in the
range $[0,1]$ (with appropriate symmetrization). The random entries follow a
uniform distribution, with no site-to-site correlation. Overall, this case
represents a rather unphysical scenario, but will be useful in the computations
that follow.
\item \textbf{Random network with $1/d^3$ scaling:} $A$ contains random entries
from
a uniform distribution scaled by $1/(hd)^3$, where $h$ is the Hamming
distance between two nodes. It represents a network similar to a spin chain
where all neighbor connectivities are allowed, and there is a possible spread in
the position of the nodes from their lattice sites.
\item \textbf{Dipolar scaled regular (symmetric) network:} We consider the
network to
be regular (symmetric) in two and three dimensions. With an appropriate choice
of basis, this can be converted to a Bravais lattice. Special cases of interest
are the graphene (honeycomb) lattice and the CNT (rolled honeycomb) lattices.
\item \textbf{Dipolar scaled regular network with vacancies:} Here we consider
the
regular network above and introduce vacancies that are binomially distributed
with
parameter $p$. This approximately maps to the NV diamond system, where we
consider transport
through a P1 lattice.
\end{enumerate}

\subsection{Mathematical preliminaries}
\begin{enumerate}
\item The generalized adjacency matrix $A$ of a network consists of positive
numbers in the range $[0,1]$. The matrix is symmetric and Hermitian.
\item For the norm, we will use the Frobenius norm, which is the generalized
Euclidean norm for matrices.
\begin{equation}
\|A\| = \sqrt{\sum_{i,j}^N |a_{ij}|^2} = \sqrt{\sum_{i}^N \sigma_i^2}
\end{equation}
where $\sigma_i$ are the singular values of $A$.
\end{enumerate}

\subsection{Random network}
Consider the right triangular form (R-form) of $A$,
\renewcommand{\arraystretch}{1}
\begin{equation}
A = \left[ 
\begin{array}{cccccc}{0}&{\times}&{\times}&{\times}&{\times}&{\times}\\
{}&{0}&{\times}&{\times}&{\times}&{\times}\\
{}&{}&{0}&{\times}&{\times}&{\times}\\
{}&{}&{}&{0}&{\times}&{\times}\\
{}&{}&{}&{}&{0}&{\times}\\
{}&{}&{}&{}&{}&{0}
\end{array}
\right]
\end{equation}

Here $\times$ refers to random numbers uniformly distributed in the range
$[0,1]$.
We have $E[X^2] = \text{Var}[X] + (E[X])^2 = 1/12 + 1/4 = 1/3$. The total number
of elements in the R matrix is,
\begin{equation}
n = \sum_{j=1}^N (N-j) = \frac{N(N-1)}{2}
\end{equation}
Hence, since the random numbers are assumed to be uncorrelated from site to
site, we have,
\begin{equation}
\|A\| = \sqrt{\frac{N(N-1)}{3}} 
\label{eqn:norm-rand}
\end{equation}
\zfr{norm}(a) shows the linear scaling of the norm in \zr{norm-rand}, compared
to the numerically obtained average of 100 manifestations of random networks.
The highest eigenvalue of $A$, $E_{\text{max}}$ also scales linearly with $N$.

\subsection{Dipolar coupled random network}
Now we consider the case where there is $1/d^3$ scaling with the Hamming
distance between two nodes. This represents a network similar to a spin chain
where all neighbor connectivities are allowed, and there is a possible spread
in the position of the nodes from their lattice sites. The adjacency matrix has
the form,
\begin{equation}
A = \left[ 
\begin{array}{cccccc}{0}&{\frac{\times}{d^3}}&{\frac{\times}{(2d)^3}}&{\frac{
\times}{(3d)^3}}&{
\frac{\times}{(4d)^3}}&{\frac{\times}{(5d)^3}}\\
{}&{0}&{\frac{\times}{d^3}}&{\frac{\times}{(2d)^3}}&{\frac{\times}{(3d)^3}}&{
\frac{\times}{(4d)^3}}
\\
{}&{}&{0}&{\frac{\times}{d^3}}&{\frac{\times}{(2d)^3}}&{\frac{\times}{(3d)^3}}\\
{}&{}&{}&{0}&{\frac{\times}{d^3}}&{\frac{\times}{(2d)^3}}\\
{}&{}&{}&{}&{0}&{\frac{\times}{d^3}}\\
{}&{}&{}&{}&{}&{0}
\end{array}
\right]
\end{equation}
As before, assuming that the sites are uncorrelated for the uniform
distribution of random numbers $\times$, we have,
\begin{eqnarray}
\|A\|^2 &=& \frac{2}{3}\left[ \frac{N-1}{d^6} + \frac{N-2}{(2d)^6} +
\frac{N-3}{(3d)^6} +
\cdots +  \frac{1}{[(N-1)d]^6}\right]\\
&=&\frac{2}{3}\left[\frac{N}{d^6}\sum_{j=1}^{N-1}\frac{1}{j^6} -
\frac{1}{d^3}\sum_{j=1}^{N-1}\frac{1}{j^5}\right]
\label{eqn:e.1}
\end{eqnarray}
Consider that,
\begin{eqnarray}
\sum_{j=1}^{N-1}\frac{1}{j^6} \approx \frac{\pi^6}{945} = 1.01734\nonumber\\
\sum_{j=1}^{N-1}\frac{1}{j^5} \approx 1.036
\end{eqnarray}
and the convergence is very rapid, i.e. it is true even for small $N$. Then,

\begin{equation}
\|A\| \approx \frac{1}{d^3}\sqrt{\frac{2}{3}(N-1)}
\end{equation}

\zfr{norm}(b) shows that the $\sqrt{N}$ scaling matches very well with the
numerically obtained average norm of 100 manifestations of random dipolar
networks. The highest eigenvalue $E_{\text{max}}$ approaches a constant
$1.6/d^3$.
%%%%%%%%%%%%%a%%%%%%%%%%%%%%%%%%%%%%%%%%%%%%%%%%%%%%%%%%%%%%%%%%%%
\begin{figure}
 \centering
\includegraphics[scale=0.35]{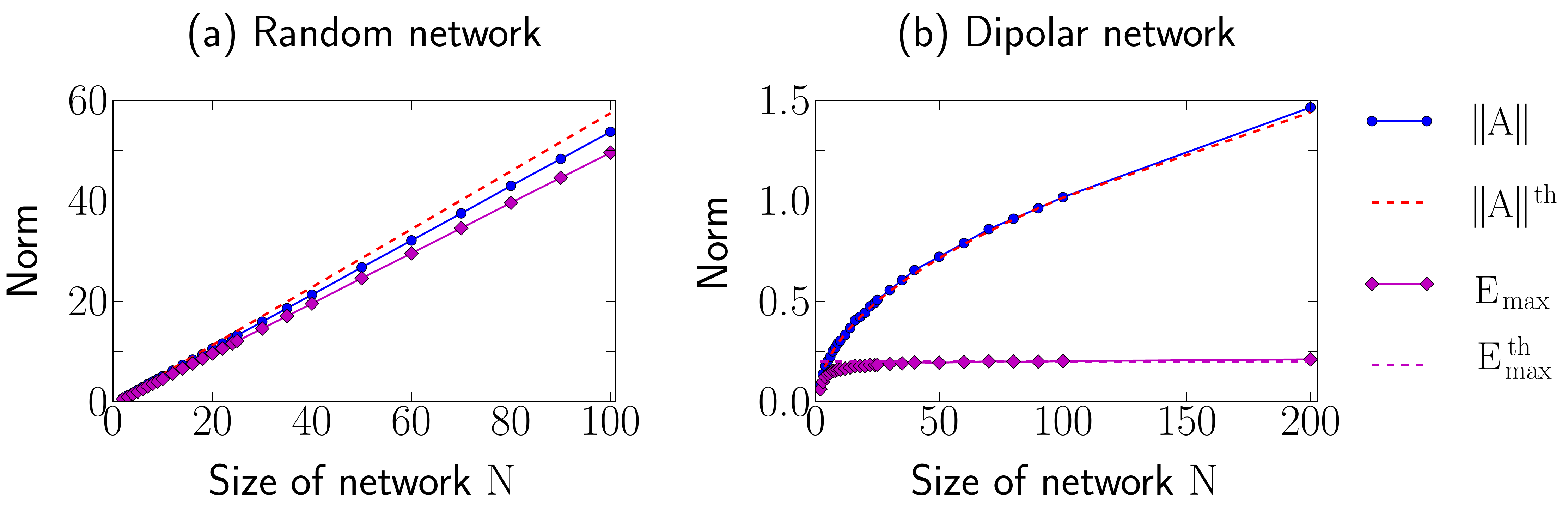}
\caption{(Color online) Figure shows the scaling with network size $N$ of the
matrix norms and largest eigenmodes of the adjacency matrices corresponding to
(a) a random network and (b) a dipolar random network. The solid lines are
average values obtained from 100 manifestations of the networks. The dashed
lines are theoretical results. For the dipolar network, the largest eigenmode
$E_{\text{max}}$ approaches a constant $1.6/d^3$ (dashed magenta line).}
\label{fig:norm}
\end{figure}
%%%%%%%%%%%%%a%%%%%%%%%%%%%%%%%%%%%%%%%%%%%%%%%%%%%%%%%%%%%%%%%%%%

\subsection{Dipolar coupled regular network}
Here we consider a regular, symmetric network in two or three dimensions. To a
good approximation, we can assume,
\begin{equation}
\|A\| = n\|A\|_{\text{cell}}
\end{equation}
where $\|A\|_{\text{cell}}$ is the adjacency matrix of the unit cell of the
underlying lattice, and $n$ is the number of tilings of this unit cell,
\begin{equation}
n\approx \frac{N}{N_{\text{cell}}}
\end{equation}
where $N_{\text{cell}}$ is the number of nodes per unit cell. 

$\|A\|_{\text{cell}}$ depends on the choice of lattice in the particular
network.
Let
us consider the case of a honeycomb lattice, where we assume only nearest
neighbor interactions. This network is found naturally in graphene and CNTs.
Then, $\|A\|_{\text{cell}} = 24/d^3$. Hence, for graphene,

\begin{equation}
\|A\| = \frac{2\sqrt{N}}{d^3}
\end{equation}

\subsection{Dipolar coupled regular network with vacancies}
Let the probability of a vacancy occurring be $p$. Once again we assume a
binomial distribution. We also assume, that we can estimate the norm in this
case by using tiling -- i.e. we consider the vacancies only in the unit cells.
Consider for simplicity the special case of graphene. For $j$ vacancies, we
have,
\begin{equation}
P_j = {}^{N_{\text{cell}}}C_{j}p^j(1-p)^{N_{\text{cell}} - j}
\end{equation}
The corresponding adjacency matrix,
\begin{equation}
\|A\|_j = \frac{4N_{\text{cell}} - j}{d^3}
\end{equation}
Hence the mean,
\begin{equation}
\|A\|_{\text{cell}} = \sum_jP_j\|A\|_j  = \frac{4N_{\text{cell}}(1-p)}{d^3}
\end{equation}
Hence,

\begin{equation}
\|A\| = \frac{2\sqrt{N(1-p)}}{d^3}
\end{equation}

\end{document}